\documentclass[twoside,twocolumn,english]{revtex4-1}
\usepackage[T1]{fontenc}
\usepackage[latin9]{inputenc}
\usepackage{geometry}
\usepackage{color}
\usepackage{amsmath}
\usepackage{amssymb}
\usepackage{enumerate}
\usepackage{graphicx}
\usepackage{mathbbol}
\usepackage{amsfonts}
\usepackage{ulem}

\makeatletter
\RequirePackage{colortbl, tabularx}
\@ifundefined{comment}{}
  {
   
  }%

\makeatother

\usepackage{babel}

\newcommand{\red}{\textcolor{red}}

\begin{document}
\global\long\def\so#1{\red{\sout{#1}}}

\global\long\def\l{\lambda}%
 
\global\long\def\ints{\mathbb{Z}}%
 
\global\long\def\nat{\mathbb{N}}%
 
\global\long\def\re{\mathbb{R}}%
 
\global\long\def\com{\mathbb{C}}%
 
\global\long\def\dff{\triangleq}%
 
\global\long\def\df{\coloneqq}%
 
\global\long\def\del{\nabla}%
 
\global\long\def\cross{\times}%
 
\global\long\def\der#1#2{\frac{d#1}{d#2}}%
 
\global\long\def\bra#1{\left\langle #1\right|}%
 
\global\long\def\ket#1{\left|#1\right\rangle }%
 
\global\long\def\braket#1#2{\left\langle #1|#2\right\rangle }%
 
\global\long\def\ketbra#1#2{\left|#1\right\rangle \left\langle #2\right|}%
 
\global\long\def\paulix{\begin{pmatrix}0  &  1\\
 1  &  0 
\end{pmatrix}}%
 
\global\long\def\pauliy{\begin{pmatrix}0  &  -i\\
 i  &  0 
\end{pmatrix}}%
 
\global\long\def\sinc{\mbox{sinc}}%
 
\global\long\def\ft{\mathcal{F}}%
 
\global\long\def\dg{\dagger}%
 
\global\long\def\bs#1{\boldsymbol{#1}}%
 
\global\long\def\norm#1{\left\Vert #1\right\Vert }%
 
\global\long\def\H{\mathcal{H}}%
 
\global\long\def\tens{\varotimes}%
 
\global\long\def\rationals{\mathbb{Q}}%
 
\global\long\def\tri{\triangle}%
 
\global\long\def\lap{\triangle}%
 
\global\long\def\e{\varepsilon}%
 
\global\long\def\broket#1#2#3{\bra{#1}#2\ket{#3}}%
 
\global\long\def\dv{\del\cdot}%
 
\global\long\def\eps{\epsilon}%
 
\global\long\def\rot{\vec{\del}\cross}%
 
\global\long\def\pd#1#2{\frac{\partial#1}{\partial#2}}%
 
\global\long\def\L{\mathcal{L}}%
 
\global\long\def\inf{\infty}%
 
\global\long\def\d{\delta}%
 
\global\long\def\I{\mathbb{I}}%
 
\global\long\def\D{\Delta}%
 
\global\long\def\r{\rho}%
 
\global\long\def\hb{\hbar}%
 
\global\long\def\s{\sigma}%
 
\global\long\def\t{\tau}%
 
\global\long\def\O{\Omega}%
 
\global\long\def\a{\alpha}%
 
\global\long\def\b{\beta}%
 
\global\long\def\th{\theta}%
 
\global\long\def\l{\lambda}%

\global\long\def\Z{\mathcal{Z}}%
 
\global\long\def\z{\zeta}%
 
\global\long\def\ord#1{\mathcal{O}\left(#1\right)}%
 
\global\long\def\ua{\uparrow}%
 
\global\long\def\da{\downarrow}%
 
\global\long\def\co#1{\left[#1\right)}%
 
\global\long\def\oc#1{\left(#1\right]}%
 
\global\long\def\tr{\mbox{tr}}%
 
\global\long\def\o{\omega}%
 
\global\long\def\nab{\del}%
 
\global\long\def\p{\psi}%
 
\global\long\def\pro{\propto}%
 
\global\long\def\vf{\varphi}%
 
\global\long\def\f{\phi}%
 
\global\long\def\mark#1#2{\underset{#2}{\underbrace{#1}}}%
 
\global\long\def\markup#1#2{\overset{#2}{\overbrace{#1}}}%
 
\global\long\def\ra{\rightarrow}%
 
\global\long\def\cd{\cdot}%
 
\global\long\def\v#1{\vec{#1}}%
 
\global\long\def\fd#1#2{\frac{\d#1}{\d#2}}%
 
\global\long\def\P{\Psi}%
 
\global\long\def\dem{\overset{\mbox{!}}{=}}%
 
\global\long\def\Lam{\Lambda}%
 
\global\long\def\m{\mu}%
 
\global\long\def\n{\nu}%

\global\long\def\ul#1{\underline{#1}}%
 
\global\long\def\at#1#2{\biggl|_{#1}^{#2}}%
 
\global\long\def\lra{\leftrightarrow}%
 
\global\long\def\var{\mbox{var}}%
 
\global\long\def\E{\mathcal{E}}%
 
\global\long\def\Op#1#2#3#4#5{#1_{#4#5}^{#2#3}}%
 
\global\long\def\up#1#2{\overset{#2}{#1}}%
 
\global\long\def\down#1#2{\underset{#2}{#1}}%
 
\global\long\def\lb{\biggl[}%
 
\global\long\def\rb{\biggl]}%
 
\global\long\def\RG{\mathfrak{R}_{b}}%
 
\global\long\def\g{\gamma}%
 
\global\long\def\Ra{\Rightarrow}%
 
\global\long\def\x{\xi}%
 
\global\long\def\c{\chi}%
 
\global\long\def\res{\mbox{Res}}%
 
\global\long\def\dif{\mathbf{d}}%
 
\global\long\def\dd{\mathbf{d}}%
 
\global\long\def\grad{\vec{\del}}%

\global\long\def\mat#1#2#3#4{\left(\begin{array}{cc}
 #1  &  #2\\
 #3  &  #4 
\end{array}\right)}%
 
\global\long\def\col#1#2{\left(\begin{array}{c}
 #1\\
 #2 
\end{array}\right)}%
 
\global\long\def\sl#1{\cancel{#1}}%
 
\global\long\def\row#1#2{\left(\begin{array}{cc}
 #1  &  ,#2\end{array}\right)}%
 
\global\long\def\roww#1#2#3{\left(\begin{array}{ccc}
 #1  &  ,#2  &  ,#3\end{array}\right)}%
 
\global\long\def\rowww#1#2#3#4{\left(\begin{array}{cccc}
 #1  &  ,#2  &  ,#3  &  ,#4\end{array}\right)}%
 
\global\long\def\matt#1#2#3#4#5#6#7#8#9{\left(\begin{array}{ccc}
 #1  &  #2  &  #3\\
 #4  &  #5  &  #6\\
 #7  &  #8  &  #9 
\end{array}\right)}%
 
\global\long\def\su{\uparrow}%
 
\global\long\def\sd{\downarrow}%
 
\global\long\def\coll#1#2#3{\left(\begin{array}{c}
 #1\\
 #2\\
 #3 
\end{array}\right)}%
 
\global\long\def\h#1{\hat{#1}}%
 
\global\long\def\colll#1#2#3#4{\left(\begin{array}{c}
 #1\\
 #2\\
 #3\\
 #4 
\end{array}\right)}%
 
\global\long\def\check{\checked}%
 
\global\long\def\v#1{\vec{#1}}%
 
\global\long\def\S{\Sigma}%
 
\global\long\def\F{\Phi}%
 
\global\long\def\M{\mathcal{M}}%
 
\global\long\def\G{\Gamma}%
 
\global\long\def\im{\mbox{Im}}%
 
\global\long\def\til#1{\tilde{#1}}%
 
\global\long\def\kb{k_{B}}%
 
\global\long\def\k{\kappa}%
 
\global\long\def\ph{\phi}%
 
\global\long\def\el{\ell}%
 
\global\long\def\en{\mathcal{N}}%
 
\global\long\def\asy{\cong}%
 
\global\long\def\sbl{\biggl[}%
 
\global\long\def\sbr{\biggl]}%
 
\global\long\def\cbl{\biggl\{}%
 
\global\long\def\cbr{\biggl\}}%
 
\global\long\def\hg#1#2{\mbox{ }_{#1}F_{#2}}%
 
\global\long\def\J{\mathcal{J}}%
 
\global\long\def\diag#1{\mbox{diag}\left[#1\right]}%
 
\global\long\def\sign#1{\mbox{sgn}\left[#1\right]}%
 
\global\long\def\T{\th}%
 
\global\long\def\rp{\reals^{+}}%

\title{Phase transition in a 1d driven tracer model}

\author{Asaf Miron$^{1}$, David Mukamel$^{1}$ and Harald A Posch$^{2}$ }
\address{$\mbox{ }^{1}$Department of Physics of Complex Systems, Weizmann
Institute of Science, Rehovot 7610001, Israel}
\address{$\mbox{ }^{2}$Computational Physics Group, Faculty of Physics, University
of Vienna, Austria}
\begin{abstract}
The effect of particle overtaking on transport in a narrow channel
is studied using a 1d model of a driven tracer in a quiescent bath.
In contrast with the well-studied non-driven case, where the tracer's
long-time dynamics changes from sub-diffusive to diffusive whenever overtaking is allowed, the driven tracer is shown to exhibit a phase transition at a finite overtaking rate. The transition separates a phase in which the stationary bath density profile, as seen in the tracer's frame, is extended, as in the non-overtaking case, to a phase with a localized bath density profile. In the extended phase the tracer velocity vanishes in the thermodynamic limit while it remains finite in the localized phase.
The phase diagram of the model as well as the tracer velocity and the bath density profile
in both phases are studied, demonstrating
their distinct features. 

\end{abstract}
\maketitle

\section{Introduction \label{sec:Introduction}}

The motion of a tracer (or a tagged particle) in a bath of identical
particles confined to a narrow channel is a classical problem which
has attracted a wealth of theoretical and experimental attention over
the past several decades. For a sufficiently narrow channel, where
particles cannot overtake one another, the motion of the tracer is
rather constrained due to strong spatial and temporal correlations
between particles, which are generated by the geometrical confinement.
The resulting dynamics is famously known as single-file diffusion,
in which the tracer's mean-square displacement $\left\langle \D x\left(t\right)^{2}\right\rangle =\left\langle \left(x\left(t\right)-\left\langle x\left(t\right)\right\rangle \right)^{2}\right\rangle $
grows sub-diffusively at long time as $\sqrt{t}$, rather than linearly
as in ordinary diffusion \citep{jepsen1965dynamics,percus1974anomalous,alexander1978diffusion}.
One aspect of single-file diffusion is its evident fragility as the channel
becomes wide enough to allow particles to overtake one another. Indeed,
as numerous studies have demonstrated, as soon as overtaking becomes
possible a smooth crossover from single-file to ordinary diffusion occurs with
diffusion dominating at asymptotically long times for \textit{any}
finite overtaking rate \citep{sane2010crossover,siems2012non,kumar2015crossover,ahmadi2017diffusion}.

New developments in experimental techniques, such as microrehology
\citep{wilson2011small} and microfluidics \citep{kirby2010micro},
have recently made it possible to control and manipulate interacting
particle fluids at the scale of a single particle, usually by employing
optical or magnetic means \citep{grier2003revolution,wittbracht2010flow}.
Such experimental tools have already been used in many studies of
complex fluids with constrained dynamics such as polymer solutions
\citep{gutsche2008colloids,kruger2009diffusion,wang2016holographic},
colloids \citep{polin2008autocalibrated,dullens2011shear} and granular
systems \citep{candelier2010journey}. On the theoretical side, these
advances have raised considerable interest in geometrically constrained
systems in which the tracer is \textit{driven} by an external force,
while the bath particles are not \citep{burlatsky1992directed,burlatsky1996motion,de1997dynamics,landim1998driven,benichou1999biased,illien2013active,cividini2016exact,cividini2016correlation,kundu2016exact,benichou2018tracer}.

A broadly studied and interesting aspect of such non-equilibrium settings is their stationary behavior. In the absence of overtaking, the mean velocity $\langle v\rangle$ of a driven tracer on a finite ring of $L$ sites is known to scale as $v\sim1/L$, vanishing as $L\ra\infty$ \citep{burlatsky1992directed,burlatsky1996motion,de1997dynamics,oshanin2004biased,illien2013active}.
Moreover, the stationary bath density profile, as seen in the tracer's
reference frame, is macroscopic and extends throughout the entire
system. We henceforth refer to a phase with these properties, namely,
vanishing of the mean velocity and a macroscopic
bath density profile, as the "extended" phase. On the other hand, for
sufficiently large overtaking rates the tracer's motion is clearly
unrestricted by the bath particles, implying a finite tracer velocity and a bath density profile
which is localized around the tracer. We refer to a phase with these
properties as the "localized" phase.

An interesting question is how does the steady state of the driven system change from extended to localized as the overtaking rates are increased. Is it like in the non-driven setting, where the sub-diffusive, single-file behavior crosses over to ordinary diffusion at any overtaking rates, or does the system remain in the extended phase for some finite overtaking rates. In this case the system would exhibit a nonequilibrium phase transition to the localized phase at some finite overtaking rates.

In this paper we address this question by considering a $1d$ driven
tracer model and studying its behavior in the presence of overtaking.
Using a combination of mean-field (MF) analysis and direct numerical
simulations, we compute the characteristic properties of the model,
including the tracer velocity and the stationary bath particle density
profile. Our studies show that the model exhibits two distinct stable phases: An extended phase and a localized phase. The extended phase remarkably
persists at finite overtaking rates, changing to the localized phase
via a continuous nonequilibrium phase transition. This surprising
observation stands in contrast with the well-established equilibrium
paradigm in which \textit{any} finite overtaking rate results in a
smooth crossover from single-file sub-diffusion to plain diffusion.

The paper is organized as follows: In Section II we introduce the
model. The main results are presented in Section III. In Section
IV, the MF analysis is presented, demonstrating the existence of two
distinct phases, the extended phase and the localized phase. There we explore various properties of these phases and the transition
between them. In Section V we describe the numerical procedure used
in the numerical analysis. In Section VI concluding remarks are given.

\section{The Model\label{sec:The-Model}}

We consider a $1d$ ring of $L$ sites labeled by $\el=0,1,...,L-1$,
occupied by one tracer particle and $N$ bath particles of average
density $\overline{\r}=\frac{N}{L-1}$. The particles interact through
simple exclusion, whereby each site holds one particle at most. Bath
particles attempt to hop to vacant neighboring right or left sites
with equal rates $1$, whereas the tracer is asymmetric, attempting
to hop to the right with rate $p$ and to the left with rate $q$.
To incorporate the possibility of overtaking into our 1d model, the
tracer tries to exchange places with a bath particle occupying its
neighboring sites to the right with rate $p'$ and to the left with
rate $q'$, if a bath particle is present. The dynamics may thus be
represented as follows 
\[
10\up{\lra}101\;\;\;\text{ ; }\;\;\;20\up{\down{\rightleftarrows}q}p02\;\;\;\text{ ; }\;\;\;21\up{\down{\rightleftarrows}{q'}}{p'}12
\]
where vacant sites, bath particles and the tracer are respectively
denoted by $0,1$ and $2$ and the rates of each process are depicted
along the corresponding arrows.

The case of vanishing exchange rates $p'=q'=0$, for which no overtaking
takes place, has been extensively studied in the past. Here we are
interested in the effect of exchange processes on the properties of
the model. To this end we shall work in the tracer's reference frame,
whose position is defined to be the site $\ell=0$, and study the
bath density profile $\r_{\el}$ at sites $\el=1,2,...,L-2,L-1$.
It is worthwhile noting that this model exhibits ``particle-hole''
symmetry, meaning that it is invariant under the simultaneous transformation
of bath particles into holes (or vacancies) implying $\r_{\el}\lra1-\r_{\el}$,
together with the transformation of hop rates into exchange rates
$p\lra p'$, $q\lra q'$ (as is evident in Eqs. \eqref{eq:discrete rho equations}
and \eqref{eq:Boundary equations}). This symmetry places bath particles
and holes, as well as hop and exchange processes, on equal footing.
For example, a tracer hopping into a vacant site (or hole) can equivalently
be thought of as the tracer exchanging places with a hole. Although
this symmetry is model-specific and not universal, its signature will
nevertheless appear in the following section which announces our main
results.

\section{Main Results\label{sec:Main-Results}}

The following results, obtained by MF calculations and numerical simulations,
demonstrate the existence of both a localized phase and a robust extended
phase, which persists in the presence of exchange. A non-equilibrium
phase transition separates the extended phase, characterized by a vanishing
tracer velocity and an extended bath density profile from the localized phase characterized by a finite
tracer velocity and a localized density profile.

The model's phase diagram is most conveniently presented when the
hopping and exchange rates are rewritten as 
\begin{equation}
\begin{cases}
p=r\left(1+\delta\right)\text{ };\text{ }q=r\left(1-\delta\right)\\
p'=r'\left(1+\delta'\right)\text{ };\text{ }q'=r'\left(1-\delta'\right)
\end{cases},\label{eq: delta and delta prime}
\end{equation}
where $r$ and $r'$ are the average hopping and exchange rates, respectively,
while $r\delta$ and $r'\delta'$ are the biases, with $r,r'\ge0$
and $-1\le\delta,\delta'\le1$. Our MF analysis yields two critical
manifolds, at the mean bath densities $\overline{\rho}_{c}^{I}$ and
$\overline{\rho}_{c}^{II}$, given by
\begin{equation}
\begin{cases}
\overline{\r}_{c}^{I}=\frac{q'\left(p-q\right)}{pq'-qp'}\equiv\frac{\delta\left(1-\delta'\right)}{\delta-\delta'}\\
\overline{\r}_{c}^{II}=\frac{p'\left(p-q\right)}{pq'-qp'}\equiv\frac{\delta\left(1+\delta'\right)}{\delta-\delta'}
\end{cases}.\label{eq:rho_c-1}
\end{equation}
Since the two manifolds are independent of the average rates $r$
and $r'$, the MF phase diagram may be represented in the 3d parameter
space $\left\{ \overline{\r},\delta,\delta'\right\} $. For convenience,
and without loss of generality, we hereafter consider $\delta>0$
(or $p>q$). The MF phase diagram in the $\left(\delta',\overline{\r}\right)$
plane is depicted in Fig. 
\begin{figure}
\begin{centering}
\includegraphics[scale=0.58]{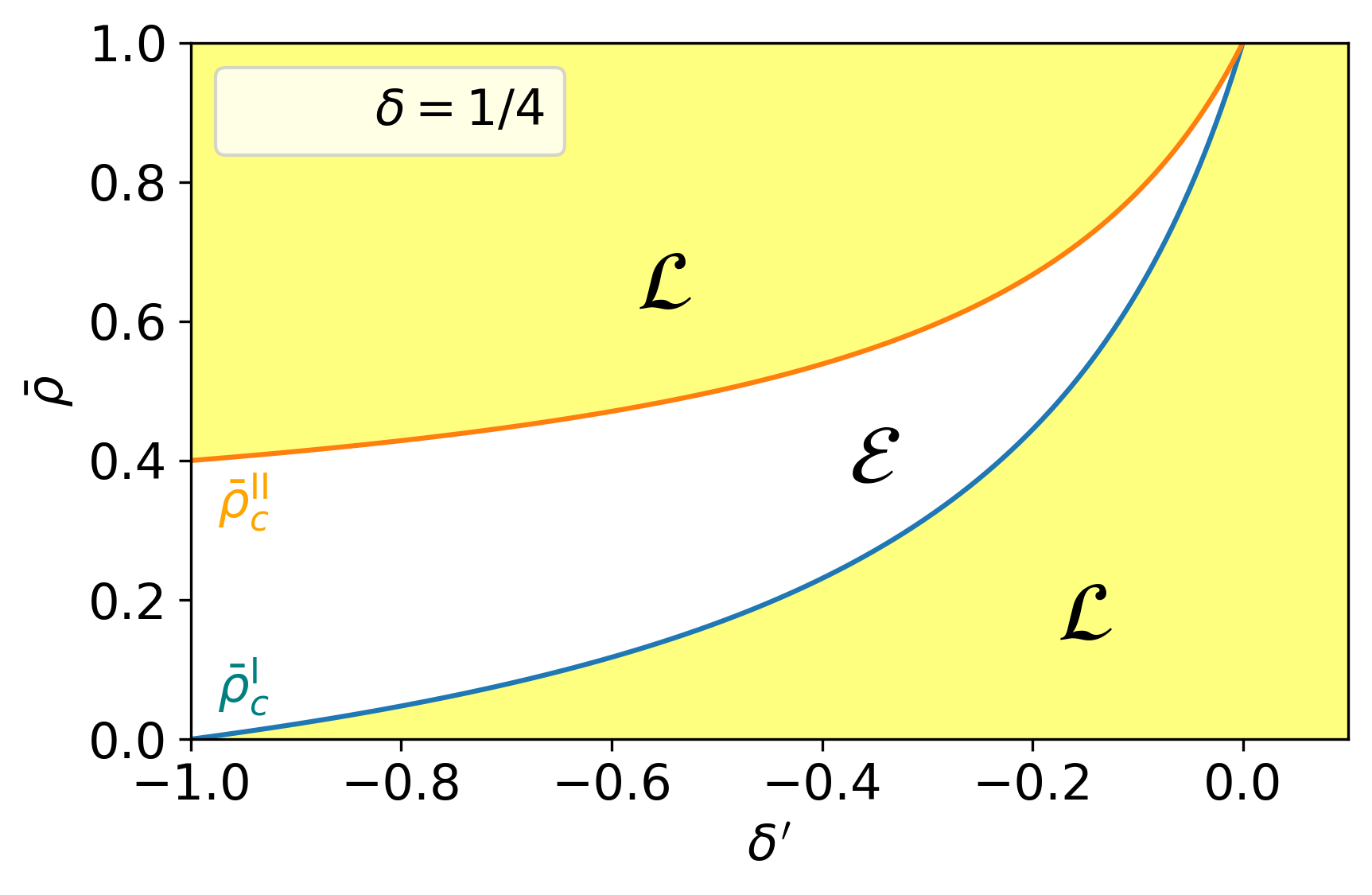} 
\par\end{centering}
\caption{The MF phase diagram in the $\left(\protect\d',\overline{\protect\r}\right)$
plane for fixed $\protect\d=3/4$, indicating the transition lines
between the localized ($\mathcal{L}$) and extended ($\mathcal{E}$) phases.}
\label{phase diagram} 
\end{figure}
\ref{phase diagram} for the hop bias $\d=3/4$. A different section,
in the $\left(\delta,\delta'\right)$ plane, is presented in Fig.
\begin{figure}
\begin{centering}
\includegraphics[scale=0.6]{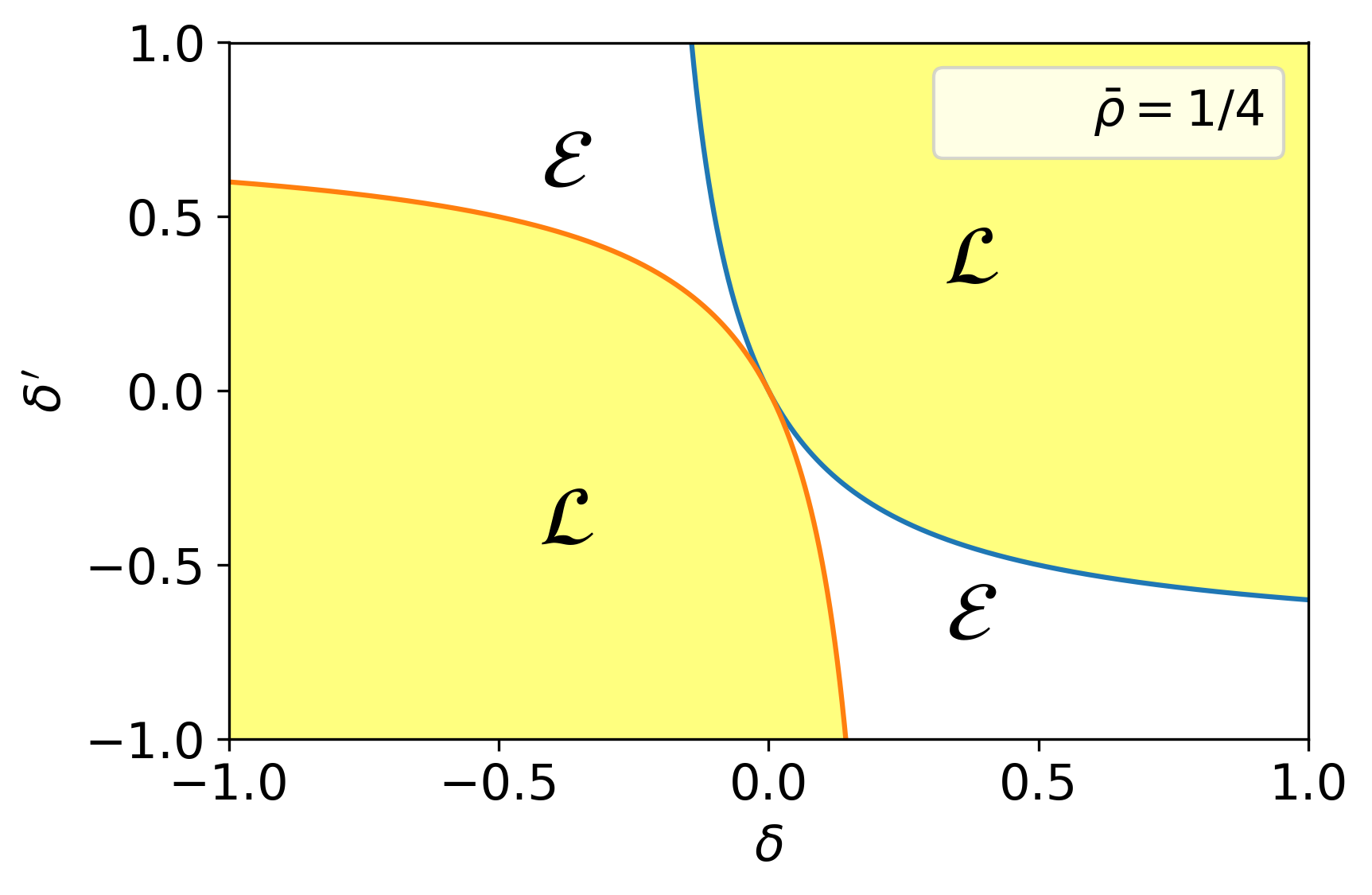} 
\par\end{centering}
\caption{The MF phase diagram for average bath density $\overline{\protect\r}=1/4$.
The localized and extended phases are respectively denoted by $\mathcal{L}$
and $\mathcal{E}$.}
\label{phase diagram_1} 
\end{figure}
\ref{phase diagram_1} for the mean density $\overline{\r}=1/4$.

The phase diagram and the nature of the phase transition may be understood
as follows: For small $\overline{\r}$ and $\delta>0$, there are
but a few bath particles behind the tracer with which it can exchange
places. Consequently, hopping dominates over exchange and a positive
stationary tracer velocity $v>0$ arises, characteristic of the localized
phase. In this phase the moving tracer generates, in its reference
frame, a stationary bath-particle density profile with an induced
density excess localized ahead of the tracer. The extent of this region
and the number of bath particles entrained in it are of $\sim\ord 1$
compared with $L$. As $\overline{\r}$ increases, more bath particles
are found behind the tracer, making the exchange process more pronounced.
For a negative exchange bias $\delta'<0$ bath particles are thus
effectively \textquotedbl pumped\textquotedbl{} from behind the tracer,
increasing the density excess region ahead. As $\overline{\r}$ approaches
the critical density $\overline{\r}_{c}^{I}$, the extent of this
region grows and the tracer's velocity $v$ continuously decreases
to zero 
\begin{equation}
v=\kappa\left(\overline{\r}_{c}^{I}-\overline{\r}\right),\label{eq:v near critical point}
\end{equation}
with the constant $\kappa$ given in Eq. \eqref{eq:v transition}.
At the transition, where $\overline{\r}=\overline{\r}_{c}^{I}$, the
density excess region ahead of the tracer becomes macroscopic and
of $\sim\ord L$, its velocity vanishes and a transition to the extended
phase takes place. The continuity of $v$ at the transition indicates
that the transition itself is continuous.

The following figures provide firm support for this picture, presenting
direct simulation results alongside MF calculations. They are generated
for the choice of hop and exchange rates $p=q'=1.75$, $q=p'=0.25$,
which conveniently correspond to the rates used in the phase diagram
in Fig. \ref{phase diagram}. For clarity, we shall refer to this
particular choice as the "canonical'' rates.
\begin{figure}
\begin{centering}
\includegraphics[scale=0.6]{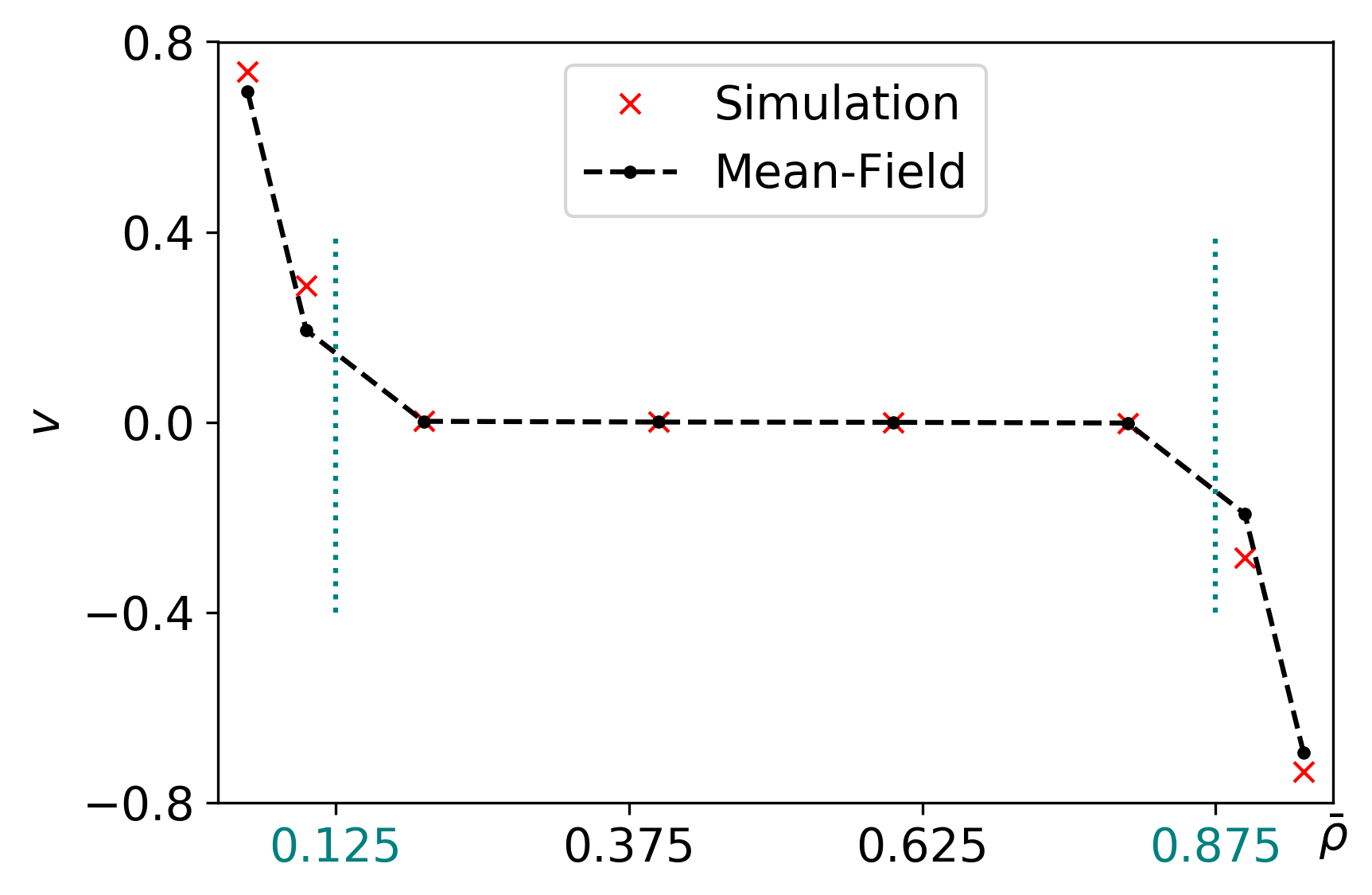} 
\par\end{centering}
\caption{The tracer velocity $v$ versus the bath density $\overline{\protect\r}$
as obtained from MF analysis and numerical simulations for $L=4096$
and the canonical rates. The vertical dotted (teal) lines indicate
the critical values predicted by MF (Eq. \eqref{eq:rho_c-1}). The
dashed black line in-between the MF values serves as a guide for the
eye.}
\label{num MF-2} 
\end{figure}

\begin{figure}
\begin{centering}
\includegraphics[scale=0.6]{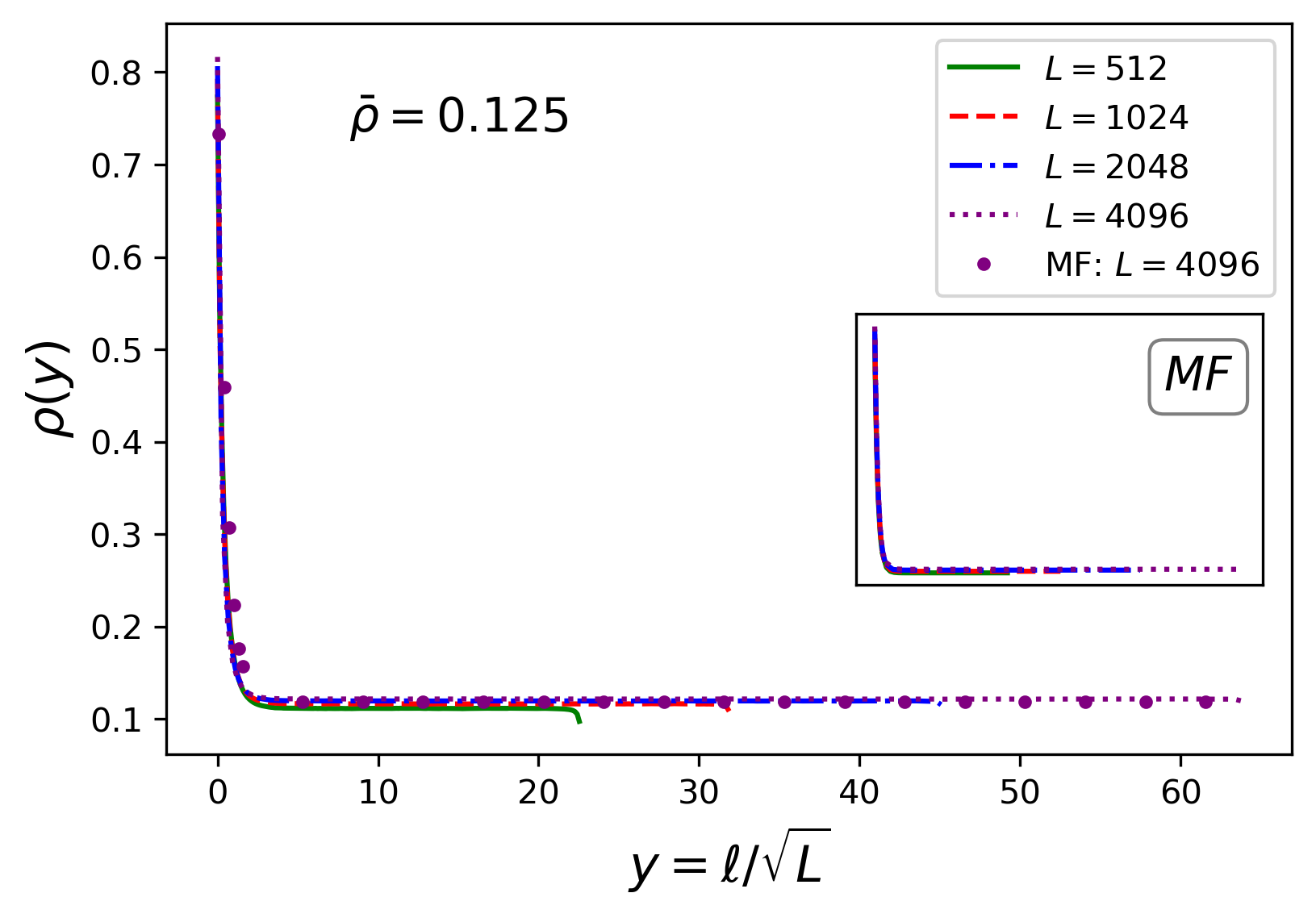} 
\par\end{centering}
\caption{Data collapse of the density profile $\protect\r\left(y\right)$ versus
$y=\protect\el/\sqrt{L}$ for different values of $L$, at the critical
density $\overline{\protect\r}=\overline{\protect\r}_{c}^{I}$. We
take the MF value of $\overline{\protect\r}_{c}^{I}$ which may slightly
differ from its exact value. For the canonical rates, this corresponds
to $\overline{\protect\r}=0.125$. The density profiles shown in the
main plot are obtained by direct numerical simulations, with the exception
of the purple dots. The latter represent the MF solution for $L=4096$
and are provided solely to demonstrate that the MF solution is in
excellent agreement with the simulated profiles, even at the critical
manifold. The profiles shown in the inset demonstrate the data collapse
of the MF solutions in Eq. \eqref{eq:SM density} for the same system
sizes as in the main plot.}
\label{sqrt L}
\end{figure}

In Fig. \ref{num MF-2} the tracer velocity $v$ is plotted as a function
of the mean density $\overline{\r}$ for a lattice of size $L=4096$.
At low density $\overline{\r}$, the system is in the localized phase
with $v>0$ due to the hopping bias $p>q$. When $\overline{\r}$
crosses $\overline{\r}_{c}^{I}$, the system exhibits a transition
to the extended phase in which $v\sim1/L$ is vanishingly small. Particle-hole
symmetry yields a similar picture as $\overline{\r}$ further grows
past $\overline{\r}_{c}^{II}$, leading the system back into the localized
phase, this time with $v<0$ due to the exchange bias $q'>p'$. The
$L$ dependence of $v$ is illustrated in Fig. 
\begin{figure}
\begin{centering}
\includegraphics[scale=0.6]{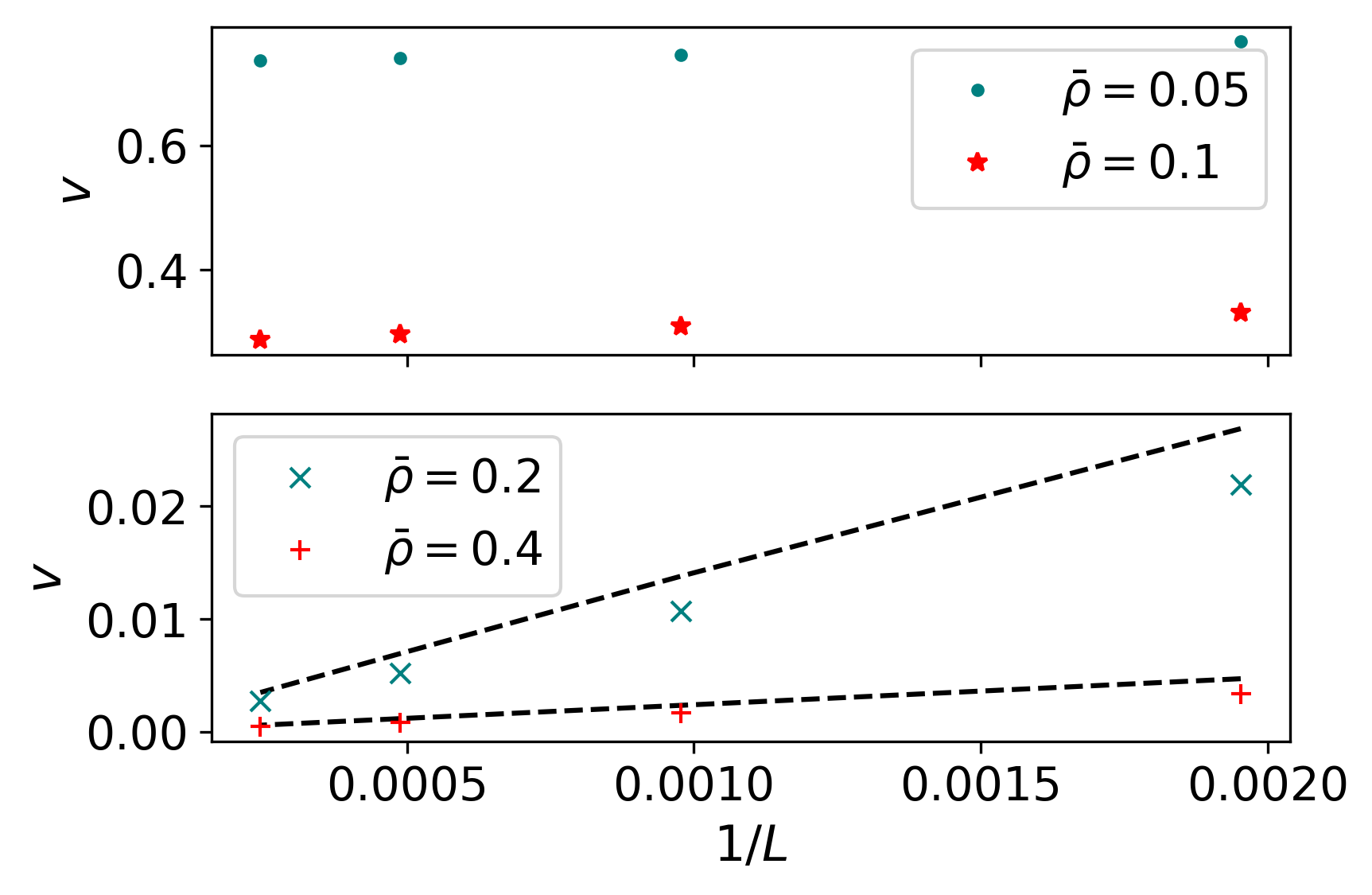} 
\par\end{centering}
\caption{The tracer velocity $v$, as obtained from numerical simulations,
is plotted against $1/L$ for different values of $\overline{\protect\r}$
in the localized and extended phases and for the canonical rates. Top Panel:
Data in the localized phase, indicating that $v$ approaches a finite
constant at large $L$. Bottom Panel: Data in the extended phase alongside
the MF solution (dashed black lines).}
\label{v} 
\end{figure}
\ref{v} in which $v$ is plotted versus the inverse system length
$1/L$. It is evident that the velocity in the extended phase vanishes as
$1/L$ whereas, in the localized phase $v$ appears to decay to a
non-zero value for large $L$. In Fig. 
\begin{figure*}
\centering{}\includegraphics[scale=0.575]{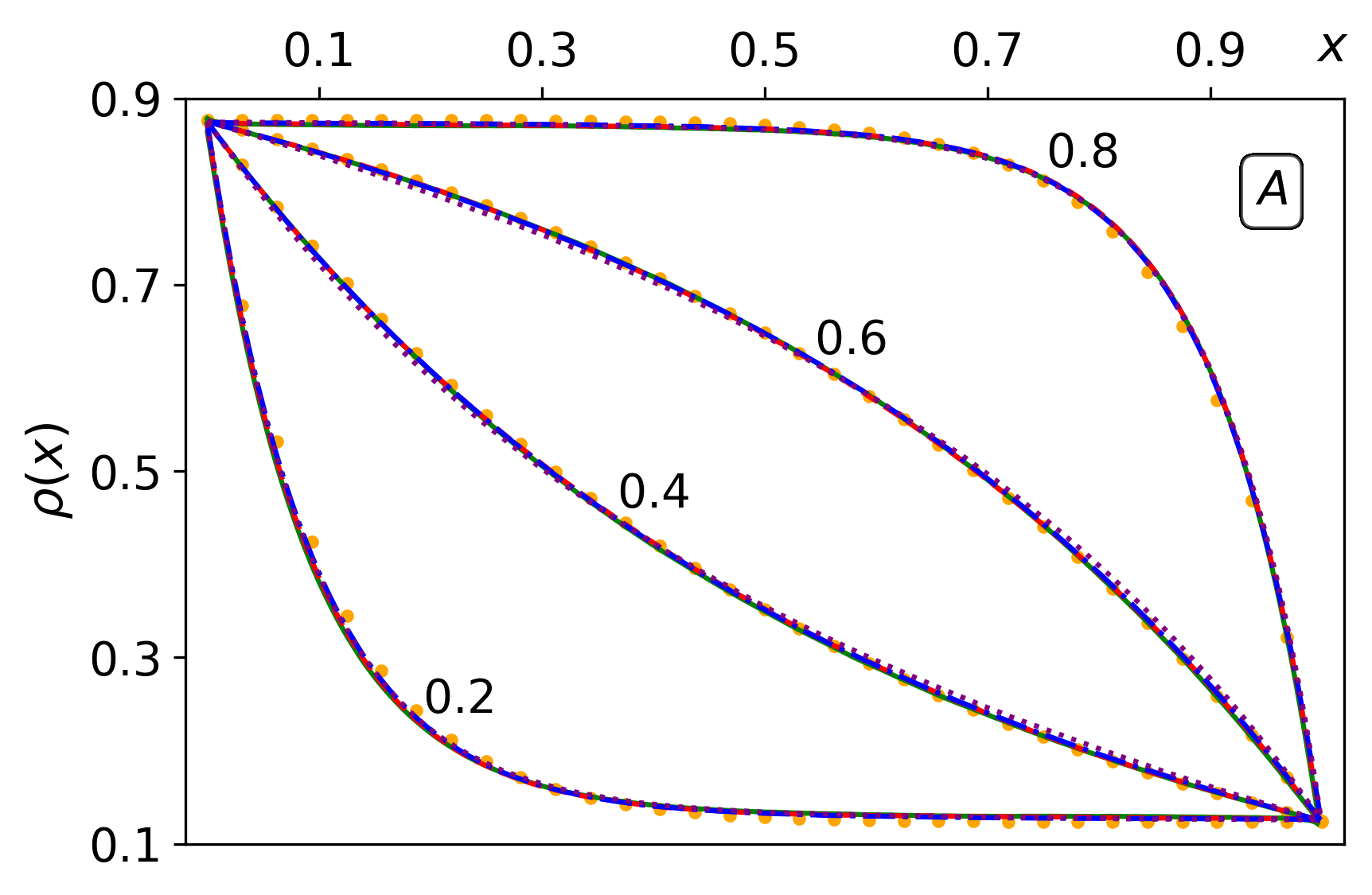}\hfill{}
\includegraphics[scale=0.6]{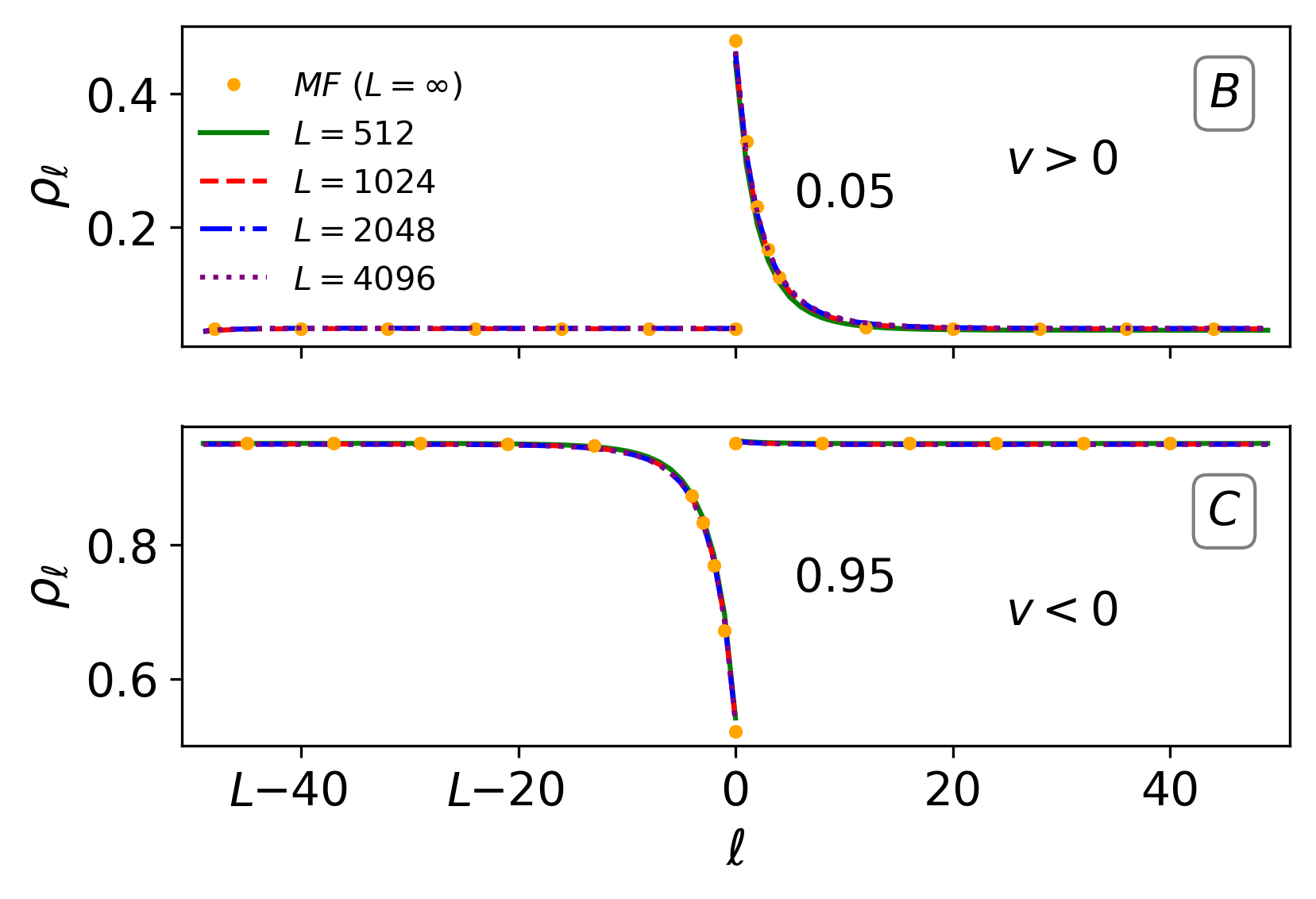}\caption{Density profile collapse, with respect to $L$, for different values
of $\overline{\protect\r}$ and the canonical rates alongside the
corresponding MF solution. The value of $\overline{\protect\r}$ is
indicated for every curve. Each panel contains the MF solution and
simulation results for four values of $L$, as listed in the figure.
Panel A describes the extended phase, where the collapse is a function of
$x=\protect\el/L$ and the density profile extends throughout the
system. Panels B and C correspond to the localized phase, where the
density profile is localized around the tracer at $\protect\el=0$.
In panel B, $v>0$ and particles accumulate to the right of the tracer
while in panel C, $v<0$ and vacancies/holes accumulate to its left.}
\label{densities} 
\end{figure*} \ref{densities} we present the stationary bath density profile in the tracer's frame. Panel A shows a data collapse as a function of
$x=\el/L$ in the extended phase, where the $\sim\ord L$ density excess
ahead of the tracer is macroscopic, extending throughout the system.
Panels B and C show a collapse as a function of $\el$ in the localized
phase. There, the $\sim\ord 1$ density excess is localized, leaving
the density in the rest of the system effectively unchanged $\sim\overline{\r}$.
The excellent agreement in Figs. \ref{num MF-2} and \ref{densities}
between the numerical simulation results and the MF calculations suggests
that the model's stationary behavior is well approximated by the MF
description.
Figure \ref{sqrt L} shows that the change from a macroscopic density
profile to a localized one takes place at the phase transition manifolds.
Mean field analysis of the density profile at the transition shows
that it spans over an intermediate distance that scales as $1/\sqrt{L}$.
Namely, that the density profile becomes a function of $y=\el/\sqrt{L}$.

\section{Mean Field Analysis \label{sec:Mean-Field-Analysis}}

We next compute the stationary properties of the model, focusing on
the bath density profile and tracer velocity. To this end, we first
formulate rate equations for the bath occupation variable $\t_{\el}\left(t\right)$
at time $t$ and sites $\el=1,2,...,L-1$ in the tracer's reference
frame, where its own position is set to $\el=0$. The occupation variable
$\t_{\el}\left(t\right)$ takes the value $1$ if site $\el$ is occupied
by a bath particle at time $t$ and $0$ otherwise. Let $\rho_{\el}(t)=\left\langle \t_{\el}\right\rangle $
be the average density at site $\el$. Applying the MF approximation
whereby correlations between site occupations are neglected, i.e.
$\left\langle \t_{k}\left(t\right)\t_{m}\left(t\right)\right\rangle \approx\r_{k}\left(t\right)\r_{m}\left(t\right)$,
the evolution equations for $\rho_{\el}(t)$ are obtained. The equation
for $\r_{\el}$ in the bulk of the system, i.e. at sites $\el\in\left[2,L-2\right]$,
is given by 
\begin{equation}
\partial_{t}\r_{\el}=\r_{\el+1}-2\r_{\el}+\r_{\el-1}+v_{+}\left(\r_{\el+1}-\r_{\el}\right)-v_{-}\left(\r_{\el}-\r_{\el-1}\right)\label{eq:discrete rho equations}
\end{equation}
whereas, at the first and last sites $\el=1$ and $\el=L-1$, the
density satisfies the boundary equations 
\begin{equation}
\begin{cases}
\partial_{t}\r_{1}=\left(1-\r_{1}\right)\left(q'\r_{L-1}+\left(1+p\right)\r_{2}\right)\\
-\left(1+p'\right)\r_{1}\left(1-\r_{2}\right)-q\r_{1}\left(1-\r_{L-1}\right) & ,\\
\partial_{t}\r_{L-1}=\left(1-\r_{L-1}\right)\left(p'\r_{1}+\left(1+q\right)\r_{L-2}\right)\\
-\left(1+q'\right)\r_{L-1}\left(1-\r_{L-2}\right)-p\r_{L-1}\left(1-\r_{1}\right).
\end{cases}\label{eq:Boundary equations}
\end{equation}
The tracer's moving rates to the right $v_{+}$ and left $v_{-}$
are given by 
\begin{equation}
v_{+}=p\left(1-\r_{1}\right)+p'\r_{1}\text{ ; }v_{-}=q\left(1-\r_{L-1}\right)+q'\r_{L-1},.\label{eq:vp,vm}
\end{equation}
The tracer velocity $v$ and total moving rate $u$ are thus given
by 
\begin{equation}
\begin{cases}
v=v_{+}-v_{-}\\
u=v_{+}+v_{-}.
\end{cases}\label{eq:v u}
\end{equation}
The stationary bulk equation \eqref{eq:discrete rho equations} then
becomes 
\begin{equation}
0=\r_{\el+1}-2\r_{\el}+\r_{\el-1}+\frac{c}{2}\left(\r_{\el+1}-\r_{\el-1}\right),\label{eq:bulk}
\end{equation}
where $c$ is 
\begin{equation}
c=v/\left(1+u/2\right).\label{eq:c}
\end{equation}
The solution of this equation is simply given by 
\begin{equation}
\r_{\el}=A+\left(\r_{1}-A\right)\left(\frac{2-c}{2+c}\right)^{\el-1},\label{eq:general profile}
\end{equation}
where the parameters $A$, $c$ and $\r_{1}$ are determined by the
boundary Eqs. \eqref{eq:Boundary equations} and by the normalization
requirement 
\begin{equation}
\sum_{\el=1}^{L-1}\r_{\el}=N.\label{eq:normalization}
\end{equation}
Within this framework, we consider two possible behaviors for the
$L$ dependence of the tracer velocity $v$, and therefor of $c$,
in the limit of large $L$. We will show that depending on the parameters
defining the model, the velocity is either $v\sim L^{-1}$ in line
with the extended phase or $v=O(1)$ as expected in the localized phase.

\subsection{The Extended Phase\label{subsec:The-Extended-Phase}}

Let us first consider the density profile of Eq. \eqref{eq:general profile} with a constant $c$ which scales at large $L$ as 
\begin{equation}
c_{\mathcal{E}}=a/L,\label{eq:c_Ex}
\end{equation}
corresponding to a tracer velocity $v_{\mathcal{E}}\sim1/L$, as expected in
the extended phase. Here $a$ is a constant which is yet to be determined.
At large $L$, a continuum limit of Eq. \eqref{eq:bulk} is obtained.
In this limit the density profile becomes a scaling function of $\el/L$
and may be expressed as $\r_{\el}^{\mathcal{E}}=\r_{\mathcal{E}}\left(x\right)$ for
the scaling variable $x\equiv\el/L$, where $x\in\left[0,1\right]$.
The solution of this equation is straightforward and given by 
\begin{equation}
\r_{\mathcal{E}}\left(x\right)=A+\left(\r_{1}^{\mathcal{E}}-A\right)e^{-ax}.\label{eq:continuum profile}
\end{equation}
The parameters $A$ and $a$ are determined by the boundary equation
\begin{equation}
\r_{L-1}^{\mathcal{E}}\approx\r_{\mathcal{E}}\left(x=1\right)=A+\left(\r_{1}^{\mathcal{E}}-A\right)e^{-a}\label{eq:continuum boundary 1}
\end{equation}
and the normalization condition $\int_{0}^{1}\dif x\r_{\mathcal{E}}\left(x\right)=\overline{\r}$,
which reduces to 
\begin{equation}
\overline{\r}=A+\left(\r_{1}^{\mathcal{E}}-A\right)W\left(a\right),\label{eq:Ex normalization}
\end{equation}
with $W\left(a\right)$ given by 
\begin{equation}
W\left(a\right)=\frac{1-e^{-a}}{a}.\label{eq:W}
\end{equation}

The densities $\r_{1}^{\mathcal{E}}$ and $\r_{L-1}^{\mathcal{E}}$ can be determined
using the boundary Eqs. \eqref{eq:Boundary equations}, which also
contains the densities $\r_{2}^{\mathcal{E}}$ and $\r_{L-2}^{\mathcal{E}}$. However,
the scaling form of the density profile with $L$ implies that $\r_{1}^{\mathcal{E}}\approx\r_{2}^{\mathcal{E}}$
and $\r_{L-1}^{\mathcal{E}}\approx\r_{L-2}^{\mathcal{E}}$, up to negligible corrections
of $\sim\ord{L^{-1}}$. As such, $\r_{1}^{\mathcal{E}}$ and $\r_{L-1}^{\mathcal{E}}$
are recovered as 
\begin{equation}
\r_{1}^{\mathcal{E}}\approx\frac{q'\left(p-q\right)}{pq'-qp'}\text{ and }\r_{L-1}^{\mathcal{E}}\approx\frac{p'\left(p-q\right)}{pq'-qp'}.\label{eq:O(1/L) rho 1 and rho L-1 sol}
\end{equation}
When inserted into Eq. \eqref{eq:continuum boundary 1} and using
the normalization condition of Eq. \eqref{eq:Ex normalization}, these
expressions give $A$ in terms of $W\left(a\right)$ and provide an
explicit (transcendental) equation for $a$, 
\begin{equation}
aW\left(a\right)\left(\r_{1}^{\mathcal{E}}-\overline{\r}\right)=\left(1-W\left(a\right)\right)\left(\r_{1}^{\mathcal{E}}-\r_{L-1}^{\mathcal{E}}\right).\label{eq:a}
\end{equation}

It is interesting to observe that the extended phase exists only when the
hopping bias $p-q$ and the exchange bias $p'-q'$ have opposite signs.
This simply follows from the normalization property of the density
profile $0\le\r_{\el}^{\mathcal{E}}\le1$ for any site $\el$ and, in particular,
$0\le\r_{1}^{\mathcal{E}},\r_{L-1}^{\mathcal{E}}\le1$. It is also important to note
that the density profile $\r_{\el}^{\mathcal{E}}$, for the particular choice
of rates $p=q'=1$ and $q=p'=0$, has previously been computed using
the matrix-product ansatz \citep{derrida2002large}. The profile obtained
by this method is exact and remarkably coincides with the MF expression
derived above. However, for this specific choice of rates, our MF
analysis shows that the extended phase persists for \uline{any value}
of the mean bath density $\overline{\r}$, implying that no phase
transition takes place for this exactly soluble case.

We conclude that the extended phase is characterized by a vanishing tracer
velocity $v_{\mathcal{E}}\pro c_{\mathcal{E}}=a/\left(L-1\right)$, with $a$ determined
by a closed transcendental equation which explicitly depends on the
rates and on the mean bath density. In addition, we find that the bath density
profile $\r_{\el}^{\mathcal{E}}$ becomes a scale function $\r_{\mathcal{E}}\left(\el/\left(L-1\right)\right)$
and extends throughout the entire system.

\subsection{The Localized Phase\label{subsec:The-Localized-Phase}}

We next consider the case of $v_{\mathcal{L}}=O(1)$ for large $L$. Since $c_{\mathcal{L}}$
remains finite at large $L$, the general form of the solution in
Eq. \eqref{eq:general profile} implies that the deviation of the
bath density profile $\r_{\el}^{\mathcal{L}}$ from the mean bath density $\overline{\r}$
must be localized to an $\sim\ord 1$ region ahead of the tracer.
In other words, $\r_{\el}^{\mathcal{L}}$ decays exponentially to $\overline{\r}$
with $\el$. For $c_{\mathcal{L}}\ge0$, the density profile is then given by
Eq. \eqref{eq:general profile} with $A=\overline{\r}$, such that
\begin{equation}
\r_{\el}^{\mathcal{L}}=\overline{\r}+\left(\r_{1}^{\mathcal{L}}-\overline{\r}\right)\left(\frac{2-c_{\mathcal{L}}}{2+c_{\mathcal{L}}}\right)^{\el-1}.\label{eq:L profile}
\end{equation}
Consequently, we find that $\r_{L-1}^{\mathcal{L}}=\r_{L-2}^{\mathcal{L}}=\overline{\r}$,
up to corrections which decay exponentially with $L$. Using these
in the boundary Eq. \eqref{eq:Boundary equations} for $\r_{L-1}^{\mathcal{L}}$,
one finds 
\begin{equation}
\r_{1}^{\mathcal{L}}=\overline{\r}\frac{p-\left(1-\overline{\r}\right)\left(q-q'\right)}{p\overline{\r}+p'\left(1-\overline{\r}\right)}.\label{eq:rho_1^L}
\end{equation}
Collecting these results allows determining $c_{\mathcal{L}}$ through the Eqs.
\eqref{eq:vp,vm}, \eqref{eq:v u} and \eqref{eq:c} as 
\begin{widetext}
\begin{equation}
c_{\mathcal{L}}=\frac{2\left[p'\left(p-q\right)-\left(pq'-qp'\right)\overline{\r}\right]}{\left(2+p\right)p'+2\left[p-p'+\left(pq+p'q'\right)\left(1-\overline{\r}\right)\right]\overline{\r}+\left[qp'\left(1-\overline{\r}\right)-pq'\overline{\r}\right]\left(1-2\overline{\r}\right)}.\label{eq:c^L}
\end{equation}
\end{widetext}

We conclude that the localized phase is characterized by a finite
tracer velocity, which explicitly depends on the rates and mean bath
density, as well as a localized bath density profile $\r_{\el}^{\mathcal{L}}$
which only deviates from $\overline{\r}$ near the tracer.

\subsection{Transition manifolds and critical behavior\label{sec:The-Critical-Manifolds}}

Let us consider the localized phase with $c_{\mathcal{L}}>0$. To determine
the transition manifolds $\overline{\r}_{c}^{I}$ and $\overline{\r}_{c}^{II}$
reported in Eq. \eqref{eq:rho_c-1} we note that $c_{\mathcal{L}}$ in Eq. \eqref{eq:c^L},
and so the tracer velocity $v_{\mathcal{L}}$ in the localized phase, vanish
at the mean bath density $\overline{\r}=\frac{p'\left(p-q\right)}{pq'-qp'}$.
Moreover, at this average density one can straightforwardly verify
(see Eqs. \eqref{eq:O(1/L) rho 1 and rho L-1 sol}, \eqref{eq:L profile},
\eqref{eq:rho_1^L}) that $\r_{1}^{\mathcal{L}}=\r_{1}^{\mathcal{E}}$ and $\r_{L-1}^{\mathcal{L}}=\r_{L-1}^{\mathcal{E}}$
implying that the density profiles in the two phases coincide. Thus,
one finds a smooth transition from the localized to the extended phase when the average density reaches $\overline{\r}_{c}^{I}=\frac{p'\left(p-q\right)}{pq'-qp'}$.
A similar analysis for $c_{\mathcal{L}}\le0$ yields the other transition manifold related to the particle-hole symmetry of the model at an average density $\overline{\r}_{c}^{II}=\frac{q'\left(p-q\right)}{pq'-qp'}$.

It is interesting to note that as the transition is approached from
the extended phase the parameter $a$ (which controls the velocity of the
tracer in this phase) diverges. This can be seen by approximating
the transcendental Eq. \eqref{eq:a} for large $a$ to obtain 
\begin{equation}
a\approx\frac{\left(p-q\right)\left(p'-q'\right)}{p'\left(p-q\right)-\left(pq'-qp'\right)\overline{\r}},\label{eq:large a}
\end{equation}
yielding a divergent $a$ at $\r=\overline{\r}_{c}^{I}$. This divergence
signifies the transition from a tracer velocity which scales like
$1/L$, in the extended phase, to the finite velocity expected in the localized phase.

The tracer's velocity $v$ may be considered as the order parameter
of the transition. It vanishes in the extended phase and it grows continuously
when the transition to the localized phase is crossed. For small deviations
from the critical density, i.e. $\overline{\r}=\overline{\r}_{c}^{I}+\d\overline{\r}$,
the velocity of the tracer in the localized phase becomes 
\begin{equation}
v_{\mathcal{L}}=-\frac{\left(p'q-q'p\right)^{2}}{pp'\left(p-q+q'-p'\right)}\delta\overline{\r}+\ord{\delta\overline{\r}^{2}}.\label{eq:v transition}
\end{equation}
In the localized phase the order parameter thus grows linearly with
the deviation of the mean density $\delta\overline{\r}$ from its
critical value $\overline{\r}_{c}^{I}$.

\subsection{Density Profile at the Transition \label{sec:The-Transition}}

In this section we study the bath density profile and the tracer velocity
at the critical manifold $\overline{\r}_{c}^{I}=\frac{p'\left(p-q\right)}{q'p-p'q}$.
A similar analysis can be carried out at the other critical manifold
$\overline{\r}_{c}^{II}$ with similar results. We show that on the
critical manifold the density profile $\r_{\el}$ varies on an intermediate
scale of $O(\sqrt{L})$ between the microscopic $O(1)$ scale of the
localized phase and the macroscopic $O(L)$ scale characterizing the
extended phase. Moreover, on this manifold the tracer's velocity scales
as $1/\sqrt{L}$.

To derive these results we consider the model at density $\overline{\r}_{c}^{I}$
and follow the analysis carries out in the extended phase. Here, though,
we take the large-$L$ continuum limit 
\begin{equation}
y=\el/L^{\a}\text{ };\text{ }c=b/L^{\a},\label{eq:SM critical scaling}
\end{equation}
with $0\le y\le L^{1-\a}$, $0\le\a\le1$ and self-consistently deduce
the values of $\a$ and $b$ at the transition. With this scaling
variables the density profile takes a form similar to that of the extended
phase 
\begin{equation}
\r\left(y\right)=A+\left(\r_{1}-A\right)e^{-by}.\label{eq:SM density}
\end{equation}
where $A,\overline{\r}_{1}$ and $b$ have to be determined by the
stationary boundary Eqs. \eqref{eq:Boundary equations} and the
normalization condition 
\begin{equation}
\overline{\r}_{c}=\frac{1}{L-1}\sum_{\el=1}^{L-1}\r_{\el}.\label{eq:SM normalziation}
\end{equation}
For large $L$ we take the continuum limit of Eq. \eqref{eq:SM normalziation},
for which 
\begin{equation}
\frac{1}{\left(L-1\right)^{\a}}\sum_{\el=1}^{L-1}\r_{\el}\longrightarrow\int_{0}^{L^{1-\a}}\dif y\r\left(y\right),\label{eq:Continuum}
\end{equation}
such that Eq. \eqref{eq:SM normalziation} yields 
\begin{equation}
A\approx\overline{\r}_{c}+\frac{\overline{\r}_{c}-\r_{1}}{bL^{1-\a}},\label{eq:SM A}
\end{equation}
up to corrections which decay exponentially with $L^{1-\a}$. Using
this result in Eq. \eqref{eq:SM density} for $\r\left(y\right)$,
one finds 
\begin{equation}
\r_{L-1}\approx\overline{\r}_{c}+\frac{\overline{\r}_{c}-\r_{1}}{bL^{1-\a}}.\label{eq:SM rho L-1}
\end{equation}

To determine $\a$, we make use of the stationary boundary Eqs. \eqref{eq:Boundary equations}
for $\r_{L-1}$. Specifically, we take 
\begin{equation}
\r_{1}=\r_{1}^{\mathcal{E}}+\d\r_{1},\label{eq:rho1 ansatz}
\end{equation}
where $\d\r_{1}$ denotes the finite-size correction to the asymptotic
(i.e. $L=\infty$) density $\r_{1}^{\mathcal{E}}=\frac{q'\left(p-q\right)}{q'p-p'q}$
at site $\el=1$ (see Eq. \eqref{eq:O(1/L) rho 1 and rho L-1 sol}).
We obtain the $L$-dependence of $\d\r_{1}$by substituting $\r_{L-1}$
of Eq. \eqref{eq:SM rho L-1} and $\r_{1}$ of Eq. \eqref{eq:rho1 ansatz}
into the boundary equation for $\r_{L-1}$, finding 
\begin{equation}
\d\r_{1}=\frac{\o}{bL^{1-\a}}+\ord{L^{2\left(\a-1\right)}},\label{eq:SM delta rho1}
\end{equation}
where 
\begin{equation}
\o=\frac{\left(p-q\right)\left(p'-q'\right)\left[p'q\left(p-q\right)-q'p\left(p'-q'\right)\right]}{pp'\left(p-q-p'+q'\right)\left(q'p-p'q\right)}.\label{eq:SM omega}
\end{equation}
Here we have used the fact that $\r_{L-2}\approx\r_{L-1}$ up to
higher order corrections.

Having obtained the $L$ dependence of $\d\r_{1}$, we finally use
it in the relation $c=b/L^{\a}$ to determine $\a$ and $b$. To this
end recall that $c=v/\left(1+u/2\right)$, where $v$ and $u$ are
given in Eqs. \eqref{eq:vp,vm} and \eqref{eq:v u}, depends explicitly
on $\r_{1}$ and $\r_{L-1}$, providing the relation 
\[
b^{2}L^{1-2\a}\approx
\]
\begin{equation}
\frac{\left(p-q\right)\left(p'-q'\right)\left(q-q'\right)-\omega\left(p-p'\right)\left(q'p-p'q\right)}{q'p\left(1+q+p'\right)-p'q\left(1+p+q'\right)},\label{eq:SM alpha relation}
\end{equation}
to leading order in $L$. Since the only $L$ dependence is on the
left hand side, we deduce that 
\begin{equation}
\a=1/2\label{eq:SM alpha}
\end{equation}
and obtain 
\begin{equation}
b^{2}\approx\frac{\left(p-q\right)\left(p'-q'\right)\left(q-q'\right)-\omega\left(p-p'\right)\left(q'p-p'q\right)}{q'p\left(1+q+p'\right)-p'q\left(1+p+q'\right)}.\label{eq:SM a^2}
\end{equation}
Combining this result for the velocity with the expressions for the
velocity in the extended and localized phases close to criticality, one
can write down the scaling form of the velocity in the vicinity of
the transition. For a density $\overline{\r}=\overline{\r}_{c}^{I}+\d\overline{\r}$,
with $\d\overline{\r}$ a small perturbation, the velocity scales
as

\begin{equation}
v\left(\d\overline{\r},L\right)=\frac{1}{\sqrt{L}}g\left(\sqrt{L}\d\overline{\r}\right),\label{eq:SM v scaling}
\end{equation}
where the scaling function $g\left(s\right)$ is given by 
\begin{equation}
g\left(s\right)=\begin{cases}
1/s & s\ra+\infty\,\,\,\,\,\,\text{(extended phase)}\\
\text{const.} & s=0\,\,\,\,\,\,\,\,\,\,\,\,\,\,\,\text{(critical)}\\
-s & s\ra-\infty\,\,\,\,\,\,\text{(localized phase)}
\end{cases}.\label{eq:SM g(x)}
\end{equation}

\section{Numerical Simulation Procedure}

In this section, the simulation procedure which has been used to obtain
the numerical results for the model introduced in Sec. \ref{sec:The-Model}
is presented.

Each realization of the dynamics began with drawing the positions
of $N$ bath particles uniformly over a lattice of $L$ sites and
then drawing the tracer position uniformly over the remaining sites.
Initial tracer hop and exchange times $\vec{\t}=\left(\t_{p},\t_{q},\t_{p'},\t_{q'}\right)$
were then drawn from exponential distributions with the respective
hop and exchange rates $p,q,p'$ and $q'$. For the bath particles,
the Gillespie algorithm was used to draw the initial bath hop time
$\s$ from an exponential distribution with rate $N/2$, accounting
for both right and left hops \citep{gillespie2007stochastic}. The
dynamics was carried out as follows: The smallest of the times $\vec{\t}$
and $\s$ was first determined. If this was $\s$, a bath particle
was drawn uniformly over the $N$ bath particle indices, as well as
a random direction $\pm1$. The bath particle would then hop to its
neighboring right/left site (in the direction $+1$ and $-1$, respectively)
if the site was vacant. If instead one of the tracer
hop times, $\t_{p}$ and $\t_{q}$, was the smallest, the tracer would
hop to the right/left neighboring site if that site was vacant. If one of the tracer exchange times, $\t_{p'}$
and $\t_{q'}$, was the smallest, the tracer would exchange places
with a bath particle to its right/left if a bath particle was present
at that site. In case the smallest time was
one of the tracer times $\vec{\t}$, the value $\pm1$ was added to
a counter which followed the position of the tracer with respect to
its initial position. For any of the above options, a corresponding
new time was drawn and the remaining times were updated. 

In Figures \ref{num MF-2}, \ref{v}, \ref{densities} and \ref{sqrt L}
which describe the stationary properties of the system, where the
relevant order of limits is $t\gg L\gg1$, the system was sampled
every sweep (consisting of $N$ hop attempts, on average) for a total
of $\sim\ord{10^{7}}$ samples (depending on the value of $L$ and
$\overline{\r}$). This number of realization has been chosen such
that no noticeable changes were detected at longer times. Each of
these figures are the result of averaging $100$ different realizations.

\section{Conclusions\label{sec:Discussion}}

In conclusion, our study suggests that geometrically-constrained driven
tracer transport may exhibit a phase transition from single-file to localized
behavior when overtaking processes are allowed. These results are based on a study of a simple lattice gas model in which the extended phase appears when the hopping bias and exchange bias are in opposing directions. This feature is particular to the model in question and may not be required to sustain the extended phase in realistic physical systems. In this context, it would be interesting to explore more realistic and detailed models of transport in geometrically constrained set-ups such as that of hard core particles moving in a narrow channel where the particle overtaking rate is controlled by the width of the channel. Molecular dynamics studies of this model will be considered separately.

The present study is focused on the steady state properties of the model. The phase transition found in this model is also expected to affect the tracer's dynamical properties, such as the temporal evolution of its mean square displacement. This is left for a future study. Another interesting direction is to study the behavior of multiple tracers in this model. Preliminary
studies show that tracers strongly attract each other, generating
a macroscopic condensate whose properties are in-line with those predicted in the extended phase. We leave this discussion to a forthcoming publication. 

\begin{acknowledgments}
We thank Julien Cividini, Satya Majumdar and Oren Raz for suggestions
and critical reading of the manuscript. This work was supported by
a research grant from the Center of Scientific Excellence at the Weizmann
Institute of Science. 
\end{acknowledgments}


\begin{thebibliography}{10}

\bibitem{jepsen1965dynamics}
DW~Jepsen.
\newblock Dynamics of a simple many-body system of hard rods.
\newblock {\em Journal of Mathematical Physics}, 6(3):405--413, 1965.

\bibitem{percus1974anomalous}
Jerome~K Percus.
\newblock Anomalous self-diffusion for one-dimensional hard cores.
\newblock {\em Physical Review A}, 9(1):557, 1974.

\bibitem{alexander1978diffusion}
S~Alexander and P~Pincus.
\newblock Diffusion of labeled particles on one-dimensional chains.
\newblock {\em Physical Review B}, 18(4):2011, 1978.

\bibitem{sane2010crossover}
Jimaan San{\'e}, Johan~T Padding, and Ard~A Louis.
\newblock The crossover from single file to fickian diffusion.
\newblock {\em Faraday discussions}, 144:285--299, 2010.

\bibitem{siems2012non}
Ullrich Siems, Christian Kreuter, Artur Erbe, Nadine Schwierz, Surajit
  Sengupta, Paul Leiderer, and Peter Nielaba.
\newblock Non-monotonic crossover from single-file to regular diffusion in
  micro-channels.
\newblock {\em Scientific reports}, 2:1015, 2012.

\bibitem{kumar2015crossover}
AV~Anil Kumar.
\newblock Crossover from normal diffusion to single-file diffusion of particles
  in a one-dimensional channel: Lj particles in zeolite zsm-22.
\newblock {\em Molecular Physics}, 113(11):1306--1310, 2015.

\bibitem{ahmadi2017diffusion}
Sheida Ahmadi and Richard~K Bowles.
\newblock Diffusion in quasi-one-dimensional channels: A small system n, p, t,
  transition state theory for hopping times.
\newblock {\em The Journal of chemical physics}, 146(15):154505, 2017.

\bibitem{wilson2011small}
Laurence~G Wilson and Wilson~CK Poon.
\newblock Small-world rheology: an introduction to probe-based active
  microrheology.
\newblock {\em Physical Chemistry Chemical Physics}, 13(22):10617--10630, 2011.

\bibitem{kirby2010micro}
Brian~J Kirby.
\newblock {\em Micro-and nanoscale fluid mechanics: transport in microfluidic
  devices}.
\newblock Cambridge university press, 2010.

\bibitem{grier2003revolution}
David~G Grier.
\newblock A revolution in optical manipulation.
\newblock {\em nature}, 424(6950):810, 2003.

\bibitem{wittbracht2010flow}
F~Wittbracht, A~Weddemann, A~Auge, and A~H{\"u}tten.
\newblock Flow guidance of magnetic particles by dipolar particle interaction.
\newblock In {\em 2010 Fourth International Conference on Quantum, Nano and
  Micro Technologies}, pages 102--106. IEEE, 2010.

\bibitem{gutsche2008colloids}
Christof Gutsche, Friedrich Kremer, Matthias Kr{\"u}ger, Markus Rauscher,
  Rudolf Weeber, and Jens Harting.
\newblock Colloids dragged through a polymer solution: Experiment, theory, and
  simulation.
\newblock {\em The Journal of chemical physics}, 129(8):084902, 2008.

\bibitem{kruger2009diffusion}
Matthias Kr{\"u}ger and Markus Rauscher.
\newblock Diffusion of a sphere in a dilute solution of polymer coils.
\newblock {\em The Journal of chemical physics}, 131(9):094902, 2009.

\bibitem{wang2016holographic}
Chen Wang, Xiao Zhong, David~B Ruffner, Alexandra Stutt, Laura~A Philips,
  Michael~D Ward, and David~G Grier.
\newblock Holographic characterization of protein aggregates.
\newblock {\em Journal of pharmaceutical sciences}, 105(3):1074--1085, 2016.

\bibitem{polin2008autocalibrated}
Marco Polin, Yohai Roichman, and David~G Grier.
\newblock Autocalibrated colloidal interaction measurements with extended
  optical traps.
\newblock {\em Physical Review E}, 77(5):051401, 2008.

\bibitem{dullens2011shear}
Roel~PA Dullens and Clemens Bechinger.
\newblock Shear thinning and local melting of colloidal crystals.
\newblock {\em Physical review letters}, 107(13):138301, 2011.

\bibitem{candelier2010journey}
Rapha{\"e}l Candelier and Olivier Dauchot.
\newblock Journey of an intruder through the fluidization and jamming
  transitions of a dense granular media.
\newblock {\em Physical Review E}, 81(1):011304, 2010.

\bibitem{burlatsky1992directed}
SF~Burlatsky, GS~Oshanin, AV~Mogutov, and M~Moreau.
\newblock Directed walk in a one-dimensional lattice gas.
\newblock {\em Physics Letters A}, 166(3-4):230--234, 1992.

\bibitem{burlatsky1996motion}
SF~Burlatsky, G~Oshanin, M~Moreau, and WP~Reinhardt.
\newblock Motion of a driven tracer particle in a one-dimensional symmetric
  lattice gas.
\newblock {\em Physical Review E}, 54(4):3165, 1996.

\bibitem{de1997dynamics}
J~De~Coninck, G~Oshanin, and M~Moreau.
\newblock Dynamics of a driven probe molecule in a liquid monolayer.
\newblock {\em EPL (Europhysics Letters)}, 38(7):527, 1997.

\bibitem{landim1998driven}
C~Landim, S~Olla, and SB~Volchan.
\newblock Driven tracer particle in one dimensional symmetric simple exclusion.
\newblock {\em Communications in mathematical physics}, 192(2):287--307, 1998.

\bibitem{benichou1999biased}
O~B{\'e}nichou, AM~Cazabat, A~Lemarchand, M~Moreau, and G~Oshanin.
\newblock Biased diffusion in a one-dimensional adsorbed monolayer.
\newblock {\em Journal of statistical physics}, 97(1-2):351--371, 1999.

\bibitem{illien2013active}
P~Illien, O~B{\'e}nichou, C~Mej{\'\i}a-Monasterio, G~Oshanin, and R~Voituriez.
\newblock Active transport in dense diffusive single-file systems.
\newblock {\em Physical review letters}, 111(3):038102, 2013.

\bibitem{cividini2016exact}
Julien Cividini, Anupam Kundu, Satya~N Majumdar, and David Mukamel.
\newblock Exact gap statistics for the random average process on a ring with a
  tracer.
\newblock {\em Journal of Physics A: Mathematical and Theoretical},
  49(8):085002, 2016.

\bibitem{cividini2016correlation}
J~Cividini, A~Kundu, Satya~N Majumdar, and D~Mukamel.
\newblock Correlation and fluctuation in a random average process on an
  infinite line with a driven tracer.
\newblock {\em Journal of Statistical Mechanics: Theory and Experiment},
  2016(5):053212, 2016.

\bibitem{kundu2016exact}
A~Kundu and J~Cividini.
\newblock Exact correlations in a single-file system with a driven tracer.
\newblock {\em EPL (Europhysics Letters)}, 115(5):54003, 2016.

\bibitem{benichou2018tracer}
O~B{\'e}nichou, P~Illien, G~Oshanin, A~Sarracino, and R~Voituriez.
\newblock Tracer diffusion in crowded narrow channels.
\newblock {\em Journal of Physics: Condensed Matter}, 30(44):443001, 2018.

\bibitem{oshanin2004biased}
G~Oshanin, O~B{\'e}nichou, SF~Burlatsky, and M~Moreau.
\newblock Biased tracer diffusion in hard-core lattice gases: Some notes on the
  validity of the einstein relation.
\newblock In {\em Instabilities and Nonequilibrium Structures IX}, pages
  33--74. Springer, 2004.

\bibitem{derrida2002large}
B~Derrida, JL~Lebowitz, and ER~Speer.
\newblock Large deviation of the density profile in the steady state of the
  open symmetric simple exclusion process.
\newblock {\em Journal of statistical physics}, 107(3-4):599--634, 2002.

\bibitem{gillespie2007stochastic}
Daniel~T Gillespie.
\newblock Stochastic simulation of chemical kinetics.
\newblock {\em Annu. Rev. Phys. Chem.}, 58:35--55, 2007.

\end{thebibliography}
\end{document}